\documentclass[conference]{IEEEtran}
\IEEEoverridecommandlockouts
\usepackage{cite}
\usepackage{amsmath,amssymb,amsfonts}
\usepackage{algorithmic}
\usepackage{graphicx}
\usepackage{textcomp}
\usepackage{hyperref}
\usepackage{tudacolors}
\usepackage{xcolor}
\usepackage{tikz}
\usepackage{pgfplots}
\def\BibTeX{{\rm B\kern-.05em{\sc i\kern-.025em b}\kern-.08em
    T\kern-.1667em\lower.7ex\hbox{E}\kern-.125emX}}
\usepackage{siunitx}

\usepackage{amsmath}
\usepackage{amssymb}
\DeclareMathOperator{\curl}{curl}
\DeclareMathOperator{\grad}{grad}
\DeclareMathOperator{\Div}{div}

\newcommand{\dd}{\,\mathrm{d}}

\definecolor{mycolor}{RGB}{146,137,233}
\usepackage{eso-pic}
\AddToShipoutPicture*{\footnotesize\sffamily\raisebox{0.75cm}{\hspace{1.65cm}\fbox{\parbox{\textwidth}{\copyright 2024 IEEE.  Personal use of this material is permitted.  Permission from IEEE must be obtained for all other uses, in any current or future media, including reprinting/republishing this material for advertising or promotional purposes, creating new collective works, for resale or redistribution to servers or lists, or reuse of any copyrighted component of this work in other works.}}}}

\begin{document}

\title{Isogeometric Shape Optimization of Multi-Tapered Coaxial Baluns Simulated by an
Integral Equation Method}

\author{\IEEEauthorblockN{Boian Balouchev}
\IEEEauthorblockA{\textit{Computational Electromagnetics}\\
\textit{Technische Universität Darmstadt}\\
Darmstadt, Germany\\
boian.balouchev@stud.tu-darmstadt.de}
\and
\IEEEauthorblockN{\hspace{1cm}}
\IEEEauthorblockA{}
\and
\IEEEauthorblockN{Jürgen Dölz}
\IEEEauthorblockA{\textit{Institute for Numerical Simulation}\\
\textit{University of Bonn}\\
Bonn, Germany\\
doelz@ins.uni-bonn.de}
\and
\IEEEauthorblockN{\hspace{1cm}}
\IEEEauthorblockA{}
\and
\IEEEauthorblockN{Maximilian Nolte}
\IEEEauthorblockA{\textit{Computational Electromagnetics}\\
\textit{Technische Universität Darmstadt}\\
Darmstadt, Germany\\
maximilian.nolte@tu-darmstadt.de}
\and
\IEEEauthorblockN{\hspace{3cm}}
\IEEEauthorblockA{}
\and
\IEEEauthorblockN{Sebastian Schöps}
\IEEEauthorblockA{\textit{Computational Electromagnetics}\\
\textit{Technische Universität Darmstadt}\\
Darmstadt, Germany\\
sebastian.schoeps@tu-darmstadt.de}
\and
\IEEEauthorblockN{\hspace{1cm}}
\IEEEauthorblockA{}
\and
\IEEEauthorblockN{Riccardo Torchio}
\IEEEauthorblockA{\textit{Department of Industrial Engineering} \\
\textit{Universita degli Studi di Padova}\\
Padova, Italy \\
riccardo.torchio@unipd.it}
\and
\IEEEauthorblockN{\hspace{3cm}}
\IEEEauthorblockA{}
}

\maketitle

\begin{abstract}
We discuss the advantages of a spline-based freeform shape optimization approach using the example of a multi-tapered coaxial balun connected to a spiral antenna.
The underlying simulation model is given in terms of a recently proposed isogeometric integral equation formulation, which can be interpreted as a high-order generalization of the partial element equivalent circuit method.
We demonstrate a significant improvement in the optimized design, i.e., a reduction in the magnitude of the scattering parameter over a wide frequency range.
\end{abstract}

\begin{IEEEkeywords}
Antenna, optimization, splines, boundary element method 
\end{IEEEkeywords}

\section{Introduction}
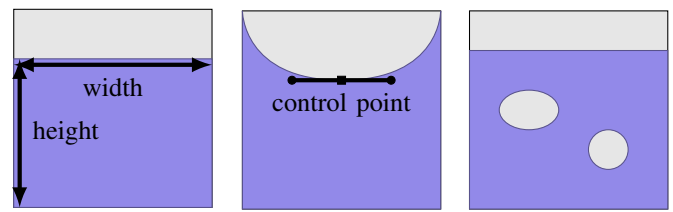
\begin{figure}[b]
    \centering
	\begin{tikzpicture}[y=0.75pt, x=0.75pt, yscale=-1, xscale=1, inner sep=0pt, outer sep=0pt]
		\draw[fill=black!10] (  0.0,  0.0) rectangle (100.0,100.0);
		\draw[fill=mycolor,draw=black!40!mycolor] (  0.0, 25.0) rectangle (100.0,100.0);
		\draw[latex-latex, ultra thick]   (  3.0, 25.0) -- (  3.0,100.0) node [midway, right=0.5em] {height};
		\draw[latex-latex, ultra thick]   (  0.0, 28.0) -- (100.0, 28.0) node [midway, below=0.5em] {width};
	\end{tikzpicture}
    ~
	\begin{tikzpicture}[y=0.75pt, x=0.75pt, yscale=-1, xscale=1, inner sep=0pt, outer sep=0pt]
		\draw[fill=black!10] (  0.0,  0.0) rectangle (100.0,100.0);
		\draw[black!40!mycolor,fill=mycolor] (  0.0,100.0) -- (  0.0,  0.0) .. controls (  0.0,0.0000) and (  0.0,35.0) .. ( 50,35.0) .. controls (100,35.0) and (100,0.0000) .. (100.0,  0.0) -- (100.0,100.0) -- cycle;
		\draw[fill=black] ( 25.0, 35.0) circle (2);
		\draw[fill=black] ( 75.0, 35.0) circle (2);
		\draw[fill=black] (48,33) rectangle (52,37);
		\draw[ultra thick,black] ( 25.0, 35.0) -- ( 75.0, 35.0) node [midway, below=0.5em] {control point};
	\end{tikzpicture}
    ~
	\begin{tikzpicture}[y=0.75pt, x=0.75pt, yscale=-1, xscale=1, inner sep=0pt, outer sep=0pt]
		\draw[fill=black!10] (  0.0,  0.0) rectangle (100.0,100.0);
		\draw[fill=mycolor, draw=black!40!mycolor] (  0.0, 20.0) rectangle (100.0,100.0);
		\draw[fill=black!10,draw=black!40!mycolor] ( 30.0, 50.0) ellipse (15.0 and 10.0);
        \draw[fill=black!10,draw=black!40!mycolor] ( 70.0, 70.0) ellipse (10.0 and 10.0);
	\end{tikzpicture}
    \caption{Parameter (left), shape (middle) and topology optimization (right).}
    \label{fig::optimization_types}
\end{figure}

In the design of electromagnetic components, the widely accepted computational alternative to a knowledge-based approach is algorithmic optimization. This topic is particularly well established in the design of antennas, see e.g., \cite{Balanis_2016aa}. The space of design parameters that defines the shape of the component is usually categorized as geometric parameters (e.g., length, width), free form shapes (e.g., represented by splines), or topologies (including holes, etc.), \autoref{fig::optimization_types}. After the design parameters are chosen, they are varied algorithmically by an optimizer, which can be broadly classified as stochastic (e.g., particle swarm) or deterministic (e.g., gradient based),~\cite{Nocedal_2006aa}.

\medskip 

Traditionally, an appropriate parameterization of the geometry must be found, and the model must be repeatedly meshed to compute the changing properties using a numerical method.
Meshing is potentially expensive and may require manual corrections, which hinders automatic approaches~\cite{Boggs_2005aa}.
In addition, the resulting designs are limited by the choice of parameters.
In contrast, freeform shape optimization greatly increases design freedom, for example, by allowing any geometric point to be moved or by acting on the underlying computer aided design (CAD) representation.
By following the paradigm of isogeometric analysis (IGA)~\cite{Cottrell_2009aa}, where the same CAD basis functions are used to represent the solution and the geometry, freeform shape optimization can be performed directly on the CAD model.%

\medskip

Furthermore, the combination of IE and IGA allows to use the boundary representation (B-rep) provided by most popular CAD software, eliminating the tedious step of volume meshing as currently proposed in~\cite{Nolte_2024aa}.
Using this formulation, we are able to optimize the triple-tapered coaxial balun proposed and analyzed in~\cite{McParland_2022aa} using a deterministic (derivative-free) optimization algorithm.
Furthermore, the IE formulation is advantageous for open
space problems, such as antenna radiation. 

\medskip

The rest of this paper is organized as follows. 
In \autoref{sec::CAD} we discuss how the computational domain is given in terms of B-splines and non-uniform rational B-splines (NURBS).
In \autoref{sec::formulation} we explain the formulation and discretization.
Then, in \autoref{sec::results} we describe the optimization procedure, we used to optimize the coaxial balun.
Finally, we summarize the results in \autoref{sec::discussion}.

\section{Boundary representation}\label{sec::CAD}
\begin{figure}[t]
    \centering
    \begin{tikzpicture}
				\begin{axis}[
					xmin=-1.2,
					xmax=5.5,
					ymin=-0.5,
					ymax=3.5,
					width=\linewidth,
					height=4.8cm,
					yticklabels={\phantom{0},\phantom{1}},
					xticklabels={\phantom{0},\phantom{1}},
					ytick style={draw=none},
					xtick style={draw=none},
				]
					\addplot[color=black,mark=*] coordinates {(1,0) (0,2) (2,3) (3,0) (4,3) (5,3) (5,1)};
					\node [black!10!mycolor, left ,font=\scriptsize] at (axis cs:1,0) {(1,0)};
					\node [black!20!mycolor, left ,font=\scriptsize] at (axis cs:0,2) {(0,2)};
					\node [black!30!mycolor, above,font=\scriptsize] at (axis cs:2,3) {(2,3)};
					\node [black!40!mycolor, below,font=\scriptsize] at (axis cs:3,0) {(3,0)};
					\node [black!50!mycolor, above,font=\scriptsize] at (axis cs:4,3) {(4,3)};
					\node [black!60!mycolor,above,font=\scriptsize] at (axis cs:5,3) {(5,3)};
					\node [black!70!mycolor, below,font=\scriptsize] at (axis cs:5,1) {(5,1)};
					\addplot[color=black,mark=x, only marks] coordinates {(2,3) (4,3)};
					\addplot[color=black,ultra thick] coordinates {(1.,0.) (0.9427,0.1191) (0.8908,0.2364) (0.8443,0.3519) (0.8032,0.4656) (0.7675,0.5775) (0.7372,0.6876) (0.7123,0.7959) (0.6928,0.9024) (0.6787,1.0071) (0.67,1.11) (0.6667,1.2111) (0.6688,1.3104) (0.6763,1.4079) (0.6892,1.5036) (0.7075,1.5975) (0.7312,1.6896) (0.7603,1.7799) (0.7948,1.8684) (0.8347,1.9551) (0.88,2.04) (0.9307,2.1231) (0.9868,2.2044) (1.0483,2.2839) (1.1152,2.3616) (1.1875,2.4375) (1.2652,2.5116) (1.3483,2.5839) (1.4368,2.6544) (1.5307,2.7231) (1.63,2.79) (1.7347,2.8551) (1.8448,2.9184) (1.9603,2.9799) (2.04,2.8824) (2.1,2.715) (2.16,2.5584) (2.22,2.4126) (2.28,2.2776) (2.34,2.1534) (2.4,2.04) (2.46,1.9374) (2.52,1.8456) (2.58,1.7646) (2.64,1.6944) (2.7,1.635) (2.76,1.5864) (2.82,1.5486) (2.88,1.5216) (2.94,1.5054) (3.,1.5) (3.06,1.5054) (3.12,1.5216) (3.18,1.5486) (3.24,1.5864) (3.3,1.635) (3.36,1.6944) (3.42,1.7646) (3.48,1.8456) (3.54,1.9374) (3.6,2.04) (3.66,2.1534) (3.72,2.2776) (3.78,2.4126) (3.84,2.5584) (3.9,2.715) (3.96,2.8824) (4.0199,2.9998) (4.0784,2.9968) (4.1351, 2.9902) (4.19,2.98) (4.2431,2.9662) (4.2944,2.9488) (4.3439, 2.9278) (4.3916,2.9032) (4.4375,2.875) (4.4816,2.8432) (4.5239,2.8078) (4.5644,2.7688) (4.6031,2.7262) (4.64,2.68) (4.6751,2.6302) (4.7084,2.5768) (4.7399,2.5198) (4.7696,2.4592) (4.7975,2.395) (4.8236,2.3272) (4.8479,2.2558) (4.8704,2.1808) (4.8911,2.1022) (4.91,2.02) (4.9271,1.9342) (4.9424,1.8448) (4.9559,1.7518) (4.9676,1.6552) (4.9775,1.555) (4.9856,1.4512) (4.9919,1.3438) (4.9964,1.2328) (4.9991,1.1182) (5.,1.)};
				\end{axis}
			\end{tikzpicture}
			\\[-0.7em]
			$\Xi=[0,0,0,1/3,1/3,1/3,2/3,2/3,2/3,1,1,1]$
			\\[0.5em]
			\begin{tikzpicture}
				\begin{axis}[
					xmin = -.05,
					xmax = 1.05,
					ymin = -.1,
					ymax = 1.1,
					xtick = {0,.33,.66,1},
					width=\linewidth,
					xticklabels = {$0$,$1/3$,$2/3$,$1$},
					ytick = {0,1},
					height=2.8cm,
					grid=major,
				]
					\addplot[black!10!mycolor,thick,mark=none,domain = 0:1/3]{1-6*x+9*x^2};
					\addplot[black!20!mycolor,thick,mark=none,domain = 0:1/3]{-6*(-x+3*x^2)};
					\addplot[black!30!mycolor,thick,mark=none,domain = 0:1/3]{9*x^2};
					\addplot[black!40!mycolor,thick,mark=none,domain = 1/3:2/3]{4-12*x+9*x^2};
					\addplot[black!50!mycolor,thick,mark=none,domain = 1/3:2/3]{-2*(2-9*x+9*x^2)};
					\addplot[black!60!mycolor,thick,mark=none,domain = 1/3:2/3]{1-6*x+9*x^2};
					\addplot[black!70!mycolor,thick,mark=none,domain = 2/3:1]{9*(1-2*x+x^2)};
					\addplot[black!80!mycolor,thick,mark=none,domain = 2/3:1]{-6*(2-5*x+3*x^2)};
					\addplot[black!90!mycolor,thick,mark=none,domain = 2/3:1]{4-12*x+9*x^2};
				\end{axis}
	\end{tikzpicture}
    \vspace{-0.5em}
    \caption{An exemplary spline curve with kinks at $x=1/3$ and $x=2/3$.}
    \label{fig::splines}
\end{figure}
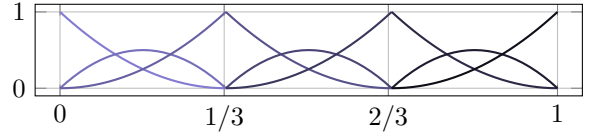

In contemporary CAD software, the computational domain is frequently given by B-rep models in terms of NURBS~\cite{Piegl_1997aa}.
The underlying basis functions are B-splines which can be defined using a $p$-open knot vector $\Xi = [\xi_1,\dots,\xi_{k + p + 1}]\in[0,1]^{k+p+1}$, where $k$ denotes the number of control points.
We can then define the basis functions for $x\in(0,1)$ and $1\leq i\leq k$ as
\begin{align}
    b_i^0(x) &= \begin{cases}
        1, \quad \mathrm{if}\;\xi_i \leq x < \xi_{i+1},\\
        0, \quad \mathrm{otherwise},
    \end{cases}
\intertext{and for any positive polynomial degree $p>0$ via the recursive formula}
    b_i^p(x) &= \frac{x - \xi_i}{\xi_{i+p} - \xi_i} b_i^{p-1}(x) +
    \frac{\xi_{i+p+1} - x}{\xi_{i+p+1} - \xi_{i+1}} b_{i+1}^{p-1}(x).
\end{align}
The B-splines $b_i^p$ span a one-dimensional function space and for higher dimensions we can obtain corresponding spaces by tensor-product construction, \cite{Hughes_2005aa}.
An example of a one-dimensional spline with two-dimensional control points that produces a piece wise smooth curve with two kinks is shown in \autoref{fig::splines} with the corresponding knot vector and basis functions.

\medskip

In applications, the computational domain is usually given by several NURBS patches, i.e.,
\begin{align}\label{eq::gamma}
\Gamma = \bigcup_{n=1}^{N_\Gamma}\Gamma_n,
\end{align}
where the parameterization of each patch $\Gamma_n$ is then given by a mapping from the two-dimensional reference domain $(0,1)\times(0,1)$ to the computational domain, via
\begin{align}\label{eq::geometry_def}
    \Gamma_n(x, y) =\!\sum_{j_1=1}^{k_1}\sum_{j_2=1}^{k_2}\frac{\pmb{p}_{j_1,j_2}^n b_{j_1}^{p_1}(x) b_{j_2}^{p_2}(y) w_{j_1,j_2}}{ \sum_{i_1=1}^{k_1}\sum_{i_2=1}^{k_2} b_{i_1}^{p_1}(x) b_{i_2}^{p_2}(y) w_{i_1,i_2}^n},
\end{align}
with three-dimensional control points $\pmb{p}_{j_1,j_2}^n$ and weights~$w_{i_1,i_2}^n$.
We assume that patches are conforming, i.e., adjacent patches share either exactly one vertex or one edge where the parameterizations coincide up to orientation.

\medskip

\section{Integral Formulation}\label{sec::formulation}
Let the free space be denoted by $\Omega$ and the perfectly electric conducting surface by $\Gamma$ as defined in \eqref{eq::gamma}.
The scattering problem is to compute for a given incident field $\pmb{E}_{\mathrm{i}}$ the scattered field $\pmb{E}_{\mathrm{s}}$ that satisfies
\begin{equation}
\left\{\begin{aligned}
    \pmb{\curl}\,\pmb{\curl} \pmb{E}_{\mathrm{s}} - \kappa^2\pmb{E}_{\mathrm{s}} &= 0\quad&&\mathrm{in}\; \Omega,\\
    \pmb{E}_{\mathrm{s}}\times\pmb{n} &= -\pmb{E}_{\mathrm{i}} \times\pmb{n}\quad&&\mathrm{on}\; \Gamma,%
\end{aligned}\right.\label{problem::maxwell}
\end{equation}
with the wave number $\kappa=\omega\sqrt{\varepsilon\mu}$ defined in terms of angular frequency $\omega$, permittivity $\varepsilon$ and permeability $\mu$ and the normal vector $\pmb n$ on $\Gamma$.
Furthermore, $\pmb{E}_{\mathrm{s}}$ needs to fulfil the Silver-Müller radiation condition~\cite[p.~29]{Kurz_2012aa}.
A solution to~\eqref{problem::maxwell} is given by the Stratton and Chu representation formula, cf., \cite[Sec.~8.14]{Stratton_1941aa}. We use the potential formulation
\begin{align}
    \pmb{E}_{\mathrm{s}}(\pmb x) = -j\omega \pmb{A}(\pmb{x}) - \pmb{\grad}\varphi(\pmb{x}),\quad\pmb{x}\in\Omega,\label{eq::EFIE}
\end{align}
with the magnetic vector potential
\begin{align}
    \pmb{A}(\pmb{x}) &= \mu\int_\Gamma \pmb{j}(\pmb{y})g_\kappa(\pmb{x},\pmb{y})\dd\Gamma_{\pmb{y}},\label{eq::BiotSavart}
\intertext{and the electric scalar potential}
    \varphi(\pmb{x}) &= \frac{1}{\varepsilon}\int_\Gamma \varrho(\pmb y)g_\kappa(\pmb{x}, \pmb{y})\dd\Gamma_{\pmb{y}}.\label{eq::CoulombIntegral}
\end{align}
Where $\pmb{j}$ is the electric surface current, $\varrho$ is the electric surface charge and 
\begin{equation}
g_\kappa(\pmb{x},\pmb{y}) = \frac{e^{-i\kappa|\pmb{x}-\pmb{y}|}}{4\pi|\pmb{x}-\pmb{y}|},
\end{equation}
denotes the homogeneous space Green's function.
We employ an A-EFIE formulation from~\cite{Qian_2008aa} with higher-order B-spline basis functions, for which a rigorous derivation can be found in~\cite{Nolte_2024aa}.
The discretization in terms of the current $\pmb{j}$ and the potential $\varphi$ leads to the linear system of equations
\begin{align}
  \begin{pmatrix}
    j\omega\pmb{L}\vphantom{\pmb{G}^{\mathrm{T}}} & \pmb{G}^{\mathrm{T}}\vphantom{\pmb{R} + j\omega\pmb{L}}
    \\ \pmb{P}\pmb{M}^{-1}\pmb{G} & -j\omega\pmb{M}
  \end{pmatrix}
  \begin{pmatrix}
    \pmb{J} \\ \pmb{\Phi}
  \end{pmatrix}
  =
  \begin{pmatrix}
    \pmb{v}_{\mathrm{ex}} \\ \pmb{0}
  \end{pmatrix}\label{eq::discretization::systemmatrix},
\end{align}
where $\pmb{J}$ and $\pmb{\Phi}$ store the degrees of freedom of $\pmb{j}$ and $\varphi$, respectively.
The matrix entries for the inductive and capacitive effects are computed as follows
\begin{align*}
    L_{ij} &= \mu\int_\Gamma \int_\Gamma g_\kappa(\pmb{x},\pmb{y}) \pmb{v}_i \cdot \pmb{v}_j \dd\Gamma_{\pmb{y}}\dd\Gamma_{\pmb{x}},\\
    P_{ij} &= \frac{1}{\varepsilon}\int_\Gamma \int_\Gamma g_\kappa(\pmb{x},\pmb{y}) v_i v_j \dd\Gamma_{\pmb{y}}\dd\Gamma_{\pmb{x}},
\end{align*}
where $\pmb{v}_i,\pmb{v}_j$ and $v_i,v_j$ are adequate vector- and scalar-valued tensor-product B-spline basis functions, respectively \cite{Nolte_2024aa,Buffa_2019ac}. The entries in the matrix blocks $\pmb{L}$ and $\pmb{P}$ in \eqref{eq::discretization::systemmatrix} contain singular integrals.
We use the Duffy trick for the quadrature of these singular integrals~\cite{Duffy_1982aa}.
For the non-singular integrals usual tensor-product Gauss-Legendre quadrature is applied~\cite{Harbrecht_2006aa}.
The remaining matrix and vector entries are given~by
\begin{align*}
    G_{ij} &= \int_\Gamma v_i \Div \pmb{v}_j \dd\Gamma_{\pmb{x}},\quad
    M_{ij} = \int_\Gamma v_i v_j \dd\Gamma_{\pmb{x}}
\end{align*}
and
\begin{align*}
    v_{\mathrm{ex},i} &= \int_\Gamma \left(\pmb{E}_{\mathrm i}\times \pmb{n}_{\pmb{x}}\right) \cdot \pmb{v}_i \dd\Gamma_{\pmb{x}}.
\end{align*}
When using lowest-order spline basis functions, the resulting formulation is equivalent to the partial element equivalent circuit (PEEC) method \cite{Ruehli_1974aa}. Even in this case, unlike classical (`orthogonal') PEEC, we can still work with arbitrarily curved spline geometries. Nonetheless, impedance $\pmb{Z}(\omega)$, admittance $\pmb{A}(\omega)$ and scattering matrices $\pmb{S}(\omega)$ can be computed from \eqref{eq::discretization::systemmatrix} as in the original circuit-like formulation, see \cite[Section 2.8]{Ruehli_2015aa}. Therefore, we will refer to our approach as IGA PEEC.

\medskip

Note, that all matrix entries and thus the solution of  \eqref{eq::discretization::systemmatrix} as well as the scattering parameter depend on the geometry which is given by the control points and weights of the spline surfaces $\Gamma_n$ in \eqref{eq::geometry_def}. By slight abuse of notation we denote all those dependencies by $\pmb{p}$ as in $\pmb{S}^{\pmb{p}}(\omega)$ for the scattering parameter.

\section{Optimization Results}\label{sec::results}
\begin{figure}[t]
    \centerline{\includegraphics[width=\linewidth]{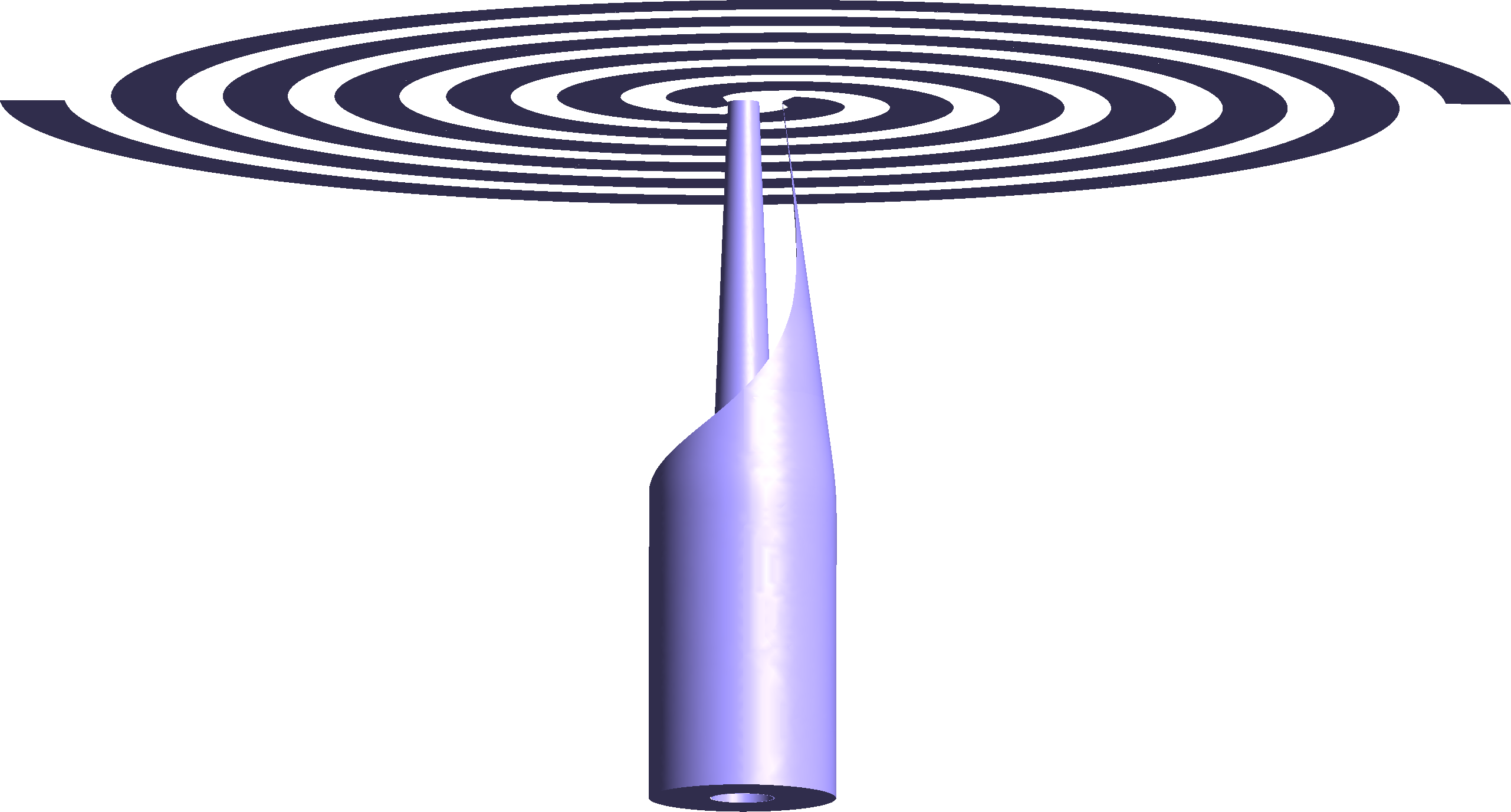}}
	\caption{Visualization of balun with attached spiral antenna with tapers based on the design of \cite{McParland_2022aa}.}
	\label{fig::balun_orig}
\end{figure}

We use the integral formulation \eqref{eq::discretization::systemmatrix} to calculate the scattering parameters for the multi-tapered coaxial balun shown in \autoref{fig::balun_orig}. The software implementation is based on Bembel~\cite{Dolz_2020ac} and uses {$N_{\Gamma}=271$} patches and {$N_\text{dof}=2871$} degrees of freedom. The multipatch NURBS model can be found at~\cite{Balouchev_2024ab}. It is motivated by the work of McParland et al., who proposed in~\cite{McParland_2022aa} a new interesting design for a coaxial balun: the unbalanced port is a coaxial cable, whose outer conductor is cut open and gradually reduced to the second wire of the balanced port.
It features a new triple-tapered design where the inner and outer conductor radius, as well as the opening angle are allowed to follow an arbitrary profile. The purpose of this structure is to convert signals between the differing ports, while also performing an impedance transformation \cite{McParland_2022aa}. One of their examples is the balun with a spiral antenna attached as shown in \autoref{fig::balun_orig}. Here, the quantity to optimize is the reflection loss over the frequency band of interest, quantified by the scattering parameter. 
McParland et al.\ optimize the design based on expert knowledge, i.e., using a combination of analytical computations, numerical simulation studies and measurements.
Notably, they suggest that more complex profiles are possible and may lead to even better results.

\medskip

We proceed to implement a freeform shape optimization process for the taper profiles.
Let $\pmb{p}$ denote the subset of the control points ($N_{\pmb{p}}=25$ points in total) defining the profiles, such that the overall geometry remains cylindrical. Furthermore, let $\mathcal{P}$ be the admissible set of control points, such that for example intersections are avoided. We solve the optimization problem
\begin{align}\label{eq::goal}
    \min_{\pmb{p} \in \mathcal{P}} \left(\sum_{i=1}^{N_\omega} |S_{11}^{\pmb{p}}(\omega_i)|^q \right)^{1/q} 
\end{align}
where $S_{11}^{\pmb{p}}(\omega_i)$ is the scattering parameter (measuring the reflection at port 1) for the geometry resulting from control points $\pmb{p}$ at angular frequency $\omega_i$. The $N_\omega=21$ frequencies $\omega_i$ are chosen as equidistant samples between $8$ and $\SI{18}{\giga \hertz}$, which corresponds to the frequency band of interest in \cite{McParland_2022aa}. The $q=8$ norm is chosen to ensure the minimization of the largest $|S_{11}^{\pmb{p}}(\omega_i)|$ values.

\medskip

\begin{figure}[t]
    \centerline{\includegraphics[width=\linewidth]{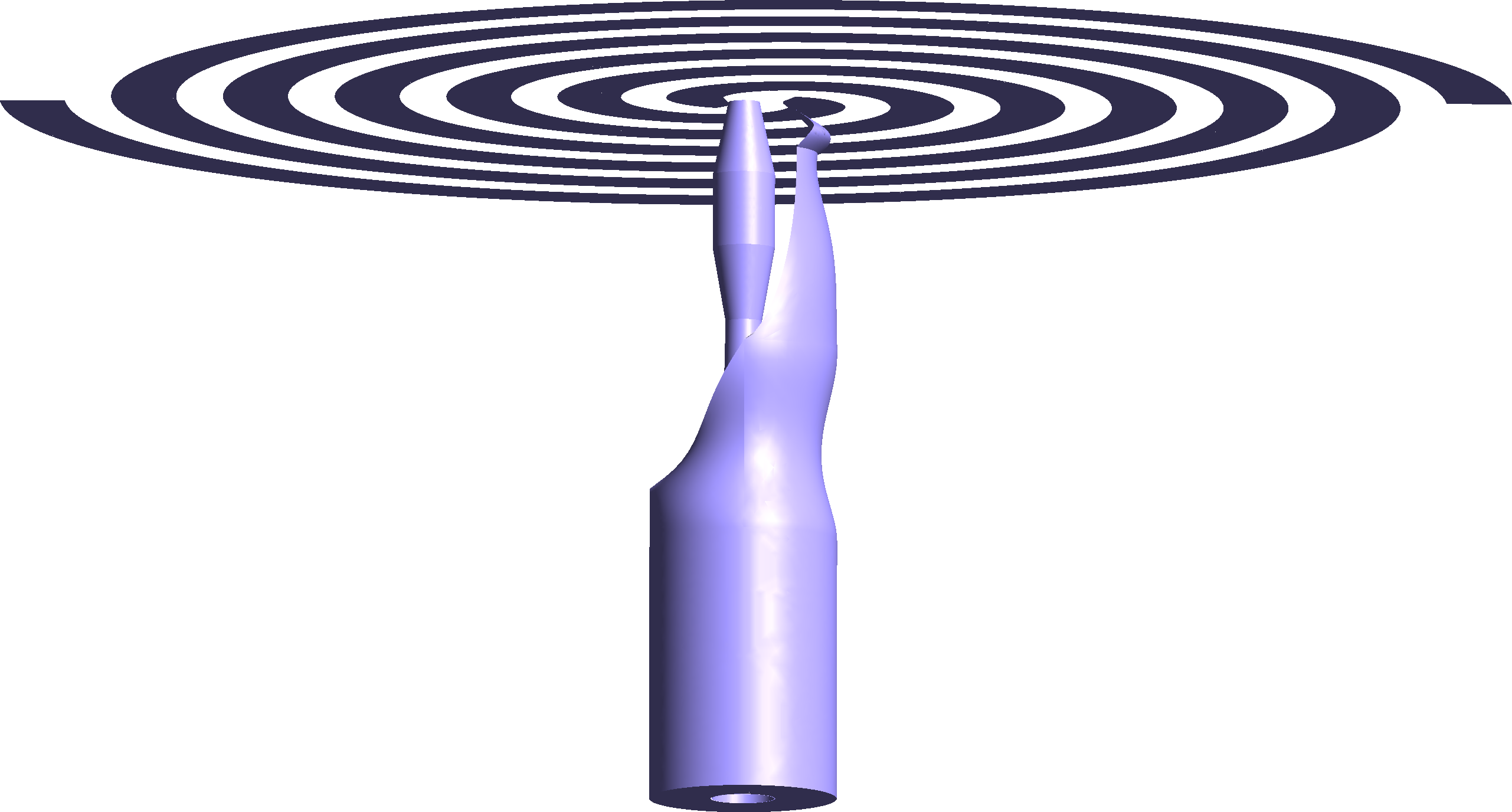}}
	\caption{Visualization of the optimized balun design.}
	\label{fig::balun_optimized}
\end{figure}

For the numerical optimization we employ the deterministic (derivative-free) bound optimization by quadratic approximation (BOBYQA) algorithm \cite{Powell_2009aa} implemented in the Python package Py-BOBYQA \cite{Cartis_2019aa,Cartis_2021aa}. We use the default parameters with global optimization and increased trust-region-radius start, end, and scaling-after-restart values ($0.2$, $0.01$, $1.2$) with which the optimizer stagnates after approx.\ $N_{\text{iter}}=80$ iterations. This process takes a few hours using significant parallelization. 

\medskip

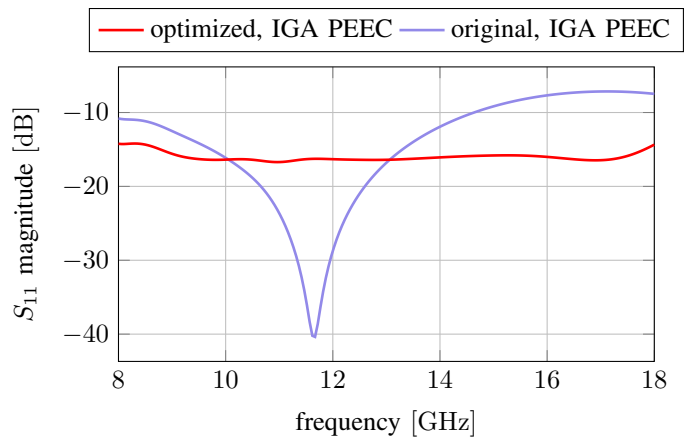
\begin{figure}[t]
    \begin{tikzpicture}
\pgfplotsset{
      every non boxed x axis/.append style={
        x axis line style={->}
      },
      every non boxed y axis/.append style={
        y axis line style={->}
      },
      every non boxed z axis/.append style={
        z axis line style={->}
      }
    }

  \begin{axis}[
    xlabel={frequency $[\si{\giga\hertz}]$},
    ylabel={$S_{11}$ magnitude $[\si{\dB}]$},
    grid=major,
    width=0.48\textwidth,
    height=5.5cm,
    xmin=8,
    xmax=18,
    reverse legend,
    legend style={at={(0.5,1.05)},anchor=south},
    legend columns = 2,
    every axis plot/.append style={line width=1pt,mark size=2pt,no markers},
  ]
  
  \addplot[mycolor] table [x=frequency,
                  y=Smag,
                  col sep=comma,
                  x expr={1e-9*\thisrow{frequency}},
                  y expr={20*log10(\thisrow{Smag})},
                  ] {data/pgfplots_our_orig.txt};
  \addlegendentry{original, IGA PEEC}
  
  \addplot table [x=frequency,
                  y=Smag,
                  col sep=comma,
                  x expr={1e-9*\thisrow{frequency}},
                  y expr={20*log10(\thisrow{Smag})},
                  ] {data/pgfplots_our_v8.txt};
  \addlegendentry{optimized, IGA PEEC}

  \end{axis}
\end{tikzpicture}
    \vspace{-2em}
 \caption{Magnitudes of scattering parameter for original and optimized geometry using $N_\omega=200$ frequency samples.}
    \label{fig::s_parameter}
\end{figure}

The reflections measured by the magnitude of $S_{11}(\omega)$ of the original design and the optimization results are visualized in \autoref{fig::s_parameter}. We observe a clear improvement, which can be quantified by a reduction of the goal function \eqref{eq::goal} by approx.\ $55$ percent where we have used $N_\omega=200$ equidistant frequency samples. To verify the results, the CAD models of the original and the optimized design are simulated with FEKO~\cite{FEKO_2023aa}. Data export and import is straightforward since the geometry data is given in the form of NURBS and can therefore be easily exported in CAD file formats such as STEP or IGES. The simulation results from FEKO are in agreement with our results and show a similar level of improvement, which confirms the effectiveness of our optimization.

\medskip

\section{Conclusion}\label{sec::discussion}
This paper demonstrates the advantages of using spline-based freeform shape optimization in the design of high-frequency components using the example of a coaxial balun connected to a spiral antenna. By using an isogeometric integral equation method, the traditional, computationally expensive re-meshing steps during the optimization procedure can be effectively bypassed.
This approach allows computation directly on CAD models, improving both geometric accuracy and computational efficiency in the optimization process.
Future research may investigate the computation and exploitation of (higher) derivatives, see e.g., \cite{Ziegler_2023ab,Ziegler_2024ab}, so that derivative-based optimization algorithms such as sequential quadratic programming (SQP) can be used, which are often more efficient~\cite{Nocedal_2006aa}.

\medskip

\section*{Acknowledgment}
The authors thank Albert E. Ruehli for the many fruitful discussions regarding PEEC and the balun example. 
This work was supported in part by the Graduate School CE within the Centre for Computational Engineering at Technische Universität Darmstadt, by the German Research Foundation under Project 443179833, the DAAD in the Framework of Short-Term Grants under 57588366 and the Future Talent Guest Stay program of TU Darmstadt. 
Finally, the support of the LOEWE project 1450/23-04 via the Hessian Ministry of Science and Art and the Hessen-Agentur is acknowledged.

\medskip
\linespread{1.11}\selectfont

\end{document}